\documentclass[showpacs,prc,preprint,nofootinbib,showkeys]{revtex4}
\pdfoutput=1

\usepackage{graphicx}
\usepackage{dcolumn}
\usepackage{bm}
\usepackage{slashed}
\usepackage{amsmath,graphicx}
\usepackage[colorlinks=true,linktocpage=true,linkcolor=blue,citecolor=blue]{hyperref}
\usepackage{float}
\usepackage{nicefrac}
\usepackage[normalem]{ulem}
\usepackage{amsmath}
\usepackage{subfigure}

\usepackage{bbold}

\usepackage[makeroom]{cancel}

\def\be{\begin{equation}}
\def\ee{\end{equation}}
\def\barr{\begin{array}}
\def\earr{\end{array}}
\def\beq{\begin{eqnarray}}
\def\eeq{\end{eqnarray}}
\def\bfig{\begin{figure}}
\def\efig{\end{figure}}
\def\lt{\left}
\def\rt{\right}

\newcommand{\p}{\partial}

\def\vv{{\boldsymbol v}}
\def\av{{\boldsymbol a}}
\def\bv{{\boldsymbol b}}
\def\xv{{\boldsymbol x}}
\def\Av{{\boldsymbol A}}
\def\Ev{{\boldsymbol E}}
\def\Bv{{\boldsymbol B}}
\def\Dv{{\boldsymbol D}}
\def\Hv{{\boldsymbol H}}
\def\Mv{{\boldsymbol M}}
\def\Pv{{\boldsymbol P}}

\def\Evc{{\boldsymbol {\cal E}}}
\def\Bvc{{\boldsymbol {\cal B}}}
\def\Dvc{{\boldsymbol {\cal D}}}
\def\Hvc{{\boldsymbol {\cal H}}}
\def\Mvc{{\boldsymbol {\cal M}}}
\def\Pvc{{\boldsymbol {\cal P}}}
\def\Jvc{{\boldsymbol {\cal J}}}

\newcommand{\rf}[1]{Eq.~(\ref{#1})}
\newcommand{\rfm}[1]{Eqs.~(\ref{#1})}
\newcommand{\rftwo}[2]{Eqs.~(\ref{#1})~and~(\ref{#2})}
\newcommand{\rfn}[1]{(\ref{#1})}

\newcommand{\nn}{\nonumber}
\newcommand{\bib}{\begin{thebibliography}}
\newcommand{\ebib}{\end{thebibliography}}
\renewcommand{\nn}{\nonumber}

\begin{document}
	
	
	\title{Vortex-like solutions and internal structures of covariant ideal magnetohydrodynamics}
	
	\author{Wojciech Florkowski}
	\email{wojciech.florkowski@ifj.edu.pl}
	\affiliation{Institute of Nuclear Physics Polish Academy of Sciences, PL-31342 Krakow, Poland}
	\affiliation{Jan Kochanowski University, PL-25406 Kielce, Poland}
	\author{Avdhesh Kumar} 
	\email{avdhesh.kumar@ifj.edu.pl} 
	\affiliation{Institute of Nuclear Physics Polish Academy of Sciences, PL-31342 Krakow, Poland}
	\author{Radoslaw Ryblewski} 
	\email{radoslaw.ryblewski@ifj.edu.pl}
	\affiliation{Institute of Nuclear Physics Polish Academy of Sciences, PL-31342 Krakow, Poland}
	\date{\today}
	\bigskip
	
	\begin{abstract}
	We discuss a manifestly covariant formulation of ideal relativistic magnetohydrodynamics, which has been recently used in astrophysical and heavy-ion contexts, and compare it to other similar frameworks. We show that the covariant equations allow for stationary vortex-like solutions that represent generalizations of the perfect-fluid solutions describing systems in global equilibrium with rotation. Such solutions are further used to demonstrate that inhomogeneous Maxwell equations, implicitly included in the covariant framework, may generate very large electric charge densities. This suggests that solutions of the covariant formulation may violate in some cases the assumptions of standard ideal magnetohydrodynamics. Furthermore, we show that the flow four-vector and conserved currents obtained in the covariant approach are usually not related to each other, which hinders kinetic-theory interpretation of the obtained results.
	\end{abstract}
	
	
	
	\maketitle 
	\newpage
\section{Introduction}\label{sec:intro}

In this paper we analyze a manifestly covariant formulation of ideal relativistic magnetohydrodynamics (MHD) defined in~Ref.~\cite{Pu:2016ayh}. This and other similar frameworks have been used recently in astrophysical~\cite{Gedalin,PhysRevE.47.4354,PhysRevE.51.4901,PhysRevLett.76.3340} and heavy-ion contexts~\cite{Florkowski:2009sw,Roy:2015kma,Inghirami:2016iru,Pang:2016yuh,Das:2017qfi,Roy:2017yvg}.  As the astrophysical applications of MHD are very broad and of established significance~\cite{Lichnerowicz}, the use of magnetohydrodynamics in heavy-ion collisions is a rather new idea, which has been triggered by the estimates \cite{Skokov:2009qp,Voronyuk:2011jd,Bzdak:2011yy} indicating that very strong magnetic fields could be present at the early stages of such processes and may affect the early dynamics  \cite{Pang:2016yuh,Das:2017qfi,Roy:2017yvg}. A related important topic is the chiral magnetic effect \cite{Kharzeev:2007jp,Fukushima:2008xe,Fukushima:2010vw}.

We refer to the MHD version of Ref.~\cite{Pu:2016ayh} as to cvMHD (covariant version). We first compare it to two more standard formulations of ideal magnetohydrodynamics~\cite{KrallTrivelpiece}. The first one does not use a manifestly covariant setup but otherwise is completely equivalent to cvMHD. The second one makes specific assumptions about the displacement vector and the system's electric charge. These two formulations are dubbed idMHD and stMHD, respectively (ideal and standard versions). The comparisons between explicitly covariant and traditional versions of MHD are useful to get more insight into physical quantities that are implicit in the covariant formulation. One of such quantities is the electric four-current. The other are the electric and magnetic fields measured in the LAB frame. 

The main aim of our work is to show that the cvMHD equations allow for stationary vortex-like solutions that are generalizations of the perfect-fluid solutions describing systems in global thermodynamic equilibrium with (a rigid) rotation~\cite{Becattini:2009wh}. The latter have become of great interest recently, since for particles with spin they suggest that fluid thermal vorticity is directly connected with the spin polarization~\cite{Becattini:2013vja}. Such identification forms now the basis for many estimates of the $\Lambda$ spin polarization in different theoretical approaches (for a recent review of this topic see, for example, Ref.~\cite{Wang:2017jpl}). However, this point of view has been recently challenged in~\cite{Florkowski:2017ruc}.
 
 Our explicit solutions of the cvMHD equations are used to demonstrate that inhomogeneous Maxwell equations (implicitly included in the cvMHD framework) may generate very large electric charge densities, in addition to large electric currents and independently of the choice done for the LAB frame~\footnote{We mean by this that there is no global reference frame where the charge density may be regarded as small in the whole space.}. This suggests that the cvMHD solutions should be a posteriori checked against possible violations of standard MHD assumptions such as the system's  quasi neutrality in a certain reference frame. Furthermore, we show that the hydrodynamic flow vector $U^\mu$, the electric current $J^\mu$, and the baryon current $N_B^\mu$, all obtained from the same covariant approach, are usually not related to each other. Therefore, the knowledge of the flow vector $U^\mu$ is not sufficient to make any conclusions about the conserved currents.  This fact also hinders kinetic-theory interpretation of the obtained results in terms of one- or two-fluid systems. We hope that our observations would be useful for correct interpretations of the physics results obtained within cvMHD.
 
Throughout the paper we use the natural units with $c=\hbar=k_B=1$. The metric tensor $g_{\mu\nu}$ has the signature $(+,-,-,-)$ and the Levi-Civita tensor $\varepsilon^{\mu\nu\alpha\beta}$ is used with the convention $\varepsilon^{0123} = -\varepsilon_{0123} = 1$. The three-vectors are marked with bold font. The scalar products of both  three- and four-vectors are denoted by a dot, hence, we may write $a \cdot b = g_{\mu\nu} a^\mu b^\nu = a^0 b^0 - \av \cdot \bv$. 

\section{MHD equations}
\label{sec:MHD}

In this section we introduce several versions of MHD, which are based on the conservation laws for energy and momentum, the Maxwell equations, and the Ohm law. The covariant formulation, which is our main topic of interest, is defined only in Sec.~\ref{sec:cvMHD}. In Secs.~\ref{sec:Tmn}--\ref{sec:stMHD} we introduce all necessary definitions and assumptions. They include the form of the energy-momentum tensor, the form of the Maxwell equations in matter, as well as the definitions of the ideal and standard MHD.

\subsection{Energy-momentum tensor}
\label{sec:Tmn}

Herein we consider a system of particles and electromagnetic (EM) fields, which is defined by the following energy-momentum tensor~\cite{Pu:2016ayh},
\be
T^{\mu\nu} = T^{\mu\nu}_\textrm{m} +  T^{\mu\nu}_\textrm{int} + T^{\mu\nu}_\textrm{em}.
\label{eq:TmnT}
\ee
Here $T^{\mu\nu}_\textrm{m}$ is the matter part~\footnote{To large extent, a division of the total energy-momentum tensor into three parts as in \rf{eq:TmnT} is arbitrary~\cite{Israel:1978up}. We note that our separation scheme differs slightly from that used in \cite{Pu:2016ayh}, however, the total $T^{\mu\nu}$ is the same. },
\beq
T^{\mu\nu}_\textrm{m} &= & \lt(\varepsilon+p\rt)\,U^\mu U^\nu - p g^{\mu\nu},
\label{eq:TmnM} 
\eeq
where $\varepsilon$ is the energy density, $p$ is the equilibrium pressure connected with $\varepsilon$ through the equation of state (EOS),  and $U^\mu$ is the fluid four-velocity. Note that the form of \rf{eq:TmnM} corresponds to the perfect-fluid description. The middle term on the right-hand side of \rf{eq:TmnT} is the interaction term,
\beq
T^{\mu\nu}_\textrm{int} &= & - \Pi^{\mu} U^{\nu} + M^{\mu\tau}F^\nu_{~\tau} ,
\label{eq:TmnI} 
\eeq
where we have introduced the Faraday electromagnetic $F_{\mu \nu}$ and magnetization $M_{\mu \nu}$  tensors. The four-vector $\Pi^{\mu}$, required for the overall consistency of the framework as argued in~\cite{Israel:1978up}, has the form~\cite{Roy:2017yvg}
\be
\Pi^{\mu} = 2 \,U_{\lambda} F^{[\mu}_{~~\nu}M^{\lambda]\nu} ,
\label{eq:Pi}
\ee
where the notation $t^{[\mu\nu]}=\frac{1}{2}\lt(t^{\mu\nu}-t^{\nu\mu}\rt)$ is used for the representation of an antisymmetric part of the rank two tensor $t^{\mu\nu}$. Below we also use the in-medium field strength tensor $H^{\mu\nu}$, which is the difference of $F_{\mu \nu}$ and $M_{\mu \nu}$,
\be
H^{\mu\nu}=F^{\mu\nu}-M^{\mu\nu}.
\label{eq:Hmn} 
\ee
The last term on the right-hand side of \rf{eq:TmnT} describes the standard EM contribution to the energy-momentum tensor
\beq
T^{\mu\nu}_\textrm{em}
&=& -F^{\mu\tau}F^\nu_{~\tau} + \frac{1}{4}g^{\mu\nu}F^{\tau\sigma}F_{\tau\sigma}.
\label{eq:TmnF}
\eeq
The energy-momentum conservation requires that the tensor \rfn{eq:TmnT} is conserved, namely
\be
\partial_{\mu}T^{\mu\nu} = 0\, .
\label{eq:TmnC}
\ee
The four equations appearing in \rf{eq:TmnC} are usually separated into one ``longitudinal'' equation (by contracting \rf{eq:TmnC} with $U_\nu$) and three ``transverse'' equations (by contracting \rf{eq:TmnC} with the projector $\Delta^{\alpha}_{\,\,\nu} = g^{\alpha}_{\,\, \nu} -U^\alpha U_\nu$). The longitudinal equation reflects the entropy conservation, while the transverse equations are relativistic versions of the hydrodynamic Euler equations.

\subsection{Fluid four-velocity and four-acceleration}
\label{sec:MEM}

The fluid four-velocity is defined as the eigen-vector of the matter part of the energy-momentum tensor, with the energy density $\varepsilon$ being the eigen-value, i.e.,  $T^{\mu}_{\textrm{m}\,\nu} U^\nu = \varepsilon U^\mu$.  The local rest frame of the fluid (LRF) is defined by the condition that $U^\mu = (1,0,0,0)$. In the LAB frame  we write $U^\mu=\gamma (1, \vv)$, where $\gamma = (1-v^2)^{-1/2}$ is the Lorentz factor and $\vv$ is the three-velocity. The four-acceleration of the fluid element is defined by the expression 
\be
A^\mu   = U^\nu \p_\nu U^\mu = \gamma \frac{dU^\mu}{dt} =  (A^0,\Av) = \gamma^4 \left( \av \cdot \vv, \av + \vv \times \left( \vv \times \av \right) \right),
\label{eq:A}
\ee
where $\av = d\vv/dt$ is the non-relativistic three-acceleration.  The four-vectors $U$ and $A$ satisfy the normalization and orthogonality conditions: $U\cdot U = 1$ and $U \cdot A =0$. We note that $A^\mu$ is defined by both $\vv$ and $\av$. The double cross product in \rf{eq:A} is a relativistic correction to $\av$  of the order $v^2/c^2$. Another relativistic correction is connected with the $\gamma^4$ factor. In the limit $v/c \to 0$ one finds that $\Av = \av$~\footnote{To study the non relativistic limit, the speed of light $c$ should be restored in \rf{eq:A}.}.

\subsection{Maxwell equations in matter}
\label{sec:MEM}

The electromagnetic field tensor $F^{\mu\nu}$ can be decomposed as follows~\cite{Gedalin,Lichnerowicz}
\be
F^{\mu\nu}=E^{\mu} U^{\nu}-E^{\nu} U^{\mu}+\epsilon^{\mu\nu\alpha\beta}U_{\alpha}B_{\beta},
\label{eq:field_strength_tensor}
\ee
with
\be
E^{\mu} = F^{\mu\nu}U_{\nu}, ~~~~~~~B^{\mu}=\frac{1}{2}\epsilon^{\mu\nu\alpha\beta}U_{\nu}F_{\alpha\beta}.
\ee
Using the above equations one can easily show that $E \cdot U = B \cdot U =0$. It can be shown also that $E \cdot E <0$ and $B \cdot B<0$, hence,  $E^{\mu}$ and $B^{\mu}$ are space-like vectors. Using this fact we can write 
\be
B^{\mu}=B \, b^{\mu}, ~~~~~~~~~~~~~~
\ee 
where, $b^{\mu}$ is a space-like unit vector which satisfies the two conditions: 
\be
b \cdot b = -1, \quad b \cdot U = 0.
\label{eq:b}
\ee

The in-medium field strength tensor $H^{\mu\nu}$ and the magnetization tensor $M^{\mu\nu}$ can be decomposed as follows
\beq
H^{\mu\nu} &=&D^{\mu}U^{\nu}-D^{\nu}U^{\mu}+\epsilon^{\mu\nu\alpha\beta} U_{\alpha}H_{\beta}, 
\label{eq:mediumresponse1} \\
M^{\mu\nu}&=&-P^{\mu}U^{\nu}+ P^{\nu}U^{\mu}+\epsilon^{\mu\nu\alpha\beta} U_{\alpha}M_{\beta},
\label{eq:mediumresponse2}
\eeq
where
\be
 D^{\mu} = H^{\mu\nu} U_{\nu}, ~~~~~~~H^{\mu}=\frac{1}{2}\epsilon^{\mu\nu\alpha\beta} U_{\nu}H_{\alpha\beta} \, ,
\ee
\be
 P^{\mu} = -M^{\mu\nu} U_{\nu}, ~~~~~~~M^{\mu}=\frac{1}{2}\epsilon^{\mu\nu\alpha\beta} U_{\nu}M_{\alpha\beta} \, .
\ee 

We emphasize, that one should distinguish between the spatial components of the four-vectors $E^\mu$ or $B^\mu$ and the electric and magnetic (three-vector) fields that usually define $F_{\mu\nu}$. For the former we use the notation $\Ev$ and $\Bv$, while for the latter we use the calligraphic bold letters $\Evc$ and $\Bvc$. Thus $E^\mu = (E^0, \Ev)$ and $B^\mu = (B^0, \Bv)$, while the EM tensor $F_{\mu\nu}$ has the components defined by the matrix
\be
F_{\mu\nu} = 
\begin{bmatrix}
0       &  \Evc^1 &  \Evc^2 &  \Evc^3 \\
- \Evc^1  &  0    & -\Bvc^3 & \Bvc^2 \\
- \Evc^2  &  \Bvc^3 & 0 & -\Bvc^1 \\
- \Evc^3  & -\Bvc^2 & \Bvc^1 & 0
\end{bmatrix}.
\ee
This means that $\Evc^i = F_{\,0i} = F^{i0}$ and $\Bvc^i = - \frac{1}{2} \epsilon^{ijk} F_{jk}$. Straightforward calculations give
\beq
E^\mu &=& (E^0, \Ev) = \gamma \left( \Evc \cdot \vv, \Evc +  \vv \times \Bvc \right), \label{calE} \\
B^\mu &=& (B^0, \Bv) = \gamma \left( \Bvc \cdot \vv, \Bvc - \vv \times \Evc \right). \label{calB} 
\eeq

Similar notation and relations can be introduced for the fields $\Mv, \Hv, \Pv, \Dv$ and $\Mvc, \Hvc, \Pvc, \Dvc$. In the local rest frame of the fluid $\Evc = \Ev$ and $\Bvc=\Bv$. The spatial parts of the four-vectors $ D^{\mu} = (D^0,\Dv)$ and $H^{\mu} = (H^0,\Hv)$ reduce in LRF to the electric displacement vector $\Dvc$ and magnetic field intensity $\Hvc$, respectively. Similarly, the spatial parts of  $P^\mu=(P^0,\Pv)$ and $M^\mu=(M^0,\Mv)$ become the electric and magnetic polarization vectors, $\Pvc$ and $\Mvc$. 
The Maxwell equations in matter are
\be
\partial_{\mu}H^{\mu\nu}=\,J^\nu
\label{eq:maxwell}
\ee
and
\be \label{eq:maxwell_dual}
\epsilon^{\mu\nu\rho\sigma}\partial_{\nu}F_{\rho\sigma}=0 \, .
\ee
We refer below to Eqs.~\rfn{eq:maxwell} and \rfn{eq:maxwell_dual} as to the inhomogeneous and homogeneous Maxwell equations, respectively.
The right-hand side of \rfm{eq:maxwell} defines the EM current $J^\mu = (\rho, \Jvc)$ that is conserved, 
\beq
\partial_\mu J^\mu = 0.
\label{eq:JC} 
\eeq
The commonly used, traditional versions of \rfm{eq:maxwell} are 
\beq
\mathbf{\nabla}\cdot \Dvc &=& \rho, \label{eq:rho1} \\
\mathbf{\nabla}\times \Hvc &=& \Jvc+ \frac{\partial \Dvc }{\partial{t}}, \label{eq:current1} 
\eeq
while for \rfm{eq:maxwell_dual} one uses the notation
\beq
\mathbf{\nabla}\cdot \Bvc&=&0, \label{eq:BGauss1} \\     
\mathbf{\nabla}\times \Evc&=&-\frac{\partial \Bvc}{\partial{t}}. \label{eq:Faraday1} 
\eeq
It is easy to see that \rftwo{eq:rho1}{eq:current1} are consistent with \rf{eq:JC}.

\subsection{Ideal MHD limit (idMHD)}
\label{sec:idMHD}

\subsubsection{The Ohm law}

In the ideal MHD  one demands that the electric field in the local rest frame vanishes, which is expressed by the condition
\be 
F^{\mu\nu} U_\nu = E^{\mu}=(E^0, \Ev)= 0. 
\label{ohm's_law}
\ee
Equation~\rfn{ohm's_law} does not imply that the electric field in other frames (in particular, in the LAB frame) is zero. In fact, from the time and space components of~\rf{calE} one gets
\be
\Evc \cdot\mathbf{v} = 0, \quad \Evc + \mathbf{v} \times \Bvc=0.
\label{ohm's_law1}
\ee
The second equation in~\rfn{ohm's_law1} means that the electric field can be obtained directly from the magnetic field and the fluid velocity. It is easy to notice that if the second equation in~\rfn{ohm's_law1} is used, the first equation in~\rfn{ohm's_law1} is automatically fulfilled.

The main argument for using \rf{ohm's_law} comes from the assumption that the EM current appearing in~\rfn{eq:JC} can be written as a sum of two terms, the convective and induced ones, namely
\beq
J^\mu = J^\mu_{\rm con} + J^\mu_{\rm ind} =  {\bar \rho} U^\mu + \sigma^{\mu\nu} E_\nu.
\label{eq:Jsum} 
\eeq
Here ${\bar \rho}$ is the LRF electric charge density and $\sigma^{\mu\nu}$ is the conductivity tensor of a system. If $\sigma^{\mu\nu}$ is very large ($\sigma^{\mu\nu} \to \infty$), we assume for consistency that $E^\mu$ is very small ($E^\mu  \to 0$) and neglect the second term~\footnote{It is also possible that the term $\sigma^{\mu\nu} E_\nu$ remains finite in this limit. We discuss this possibility below.}.

\subsubsection{Electric and magnetic polarizations}

Assuming that $P^{\mu}$ and $D^{\mu}$ are proportional to $E^{\mu}$, we conclude that they vanish as well. Note that using equations (\ref{eq:field_strength_tensor}) and (\ref{eq:mediumresponse2}) in Eq.~(\ref{eq:Pi}) we find 
\be
\Pi^{\mu} = 0,
\label{eq:Pi1}
\ee
hence, $\Pi^\mu$ can be neglected in \rf{eq:TmnI}.  Assuming further that the magnetic polarization is linearly proportional to the magnetic field, we can write
\beq
M^{\mu} = \chi_{m} B^{\mu} =  \chi_{m} B b^{\mu} \equiv M b^{\mu}.
\label{MchiB}
\eeq

In the following we assume that the susceptibility $\chi_m$ is constant. Equation \rfn{MchiB} and the fact that $E^\mu = P^\mu = D^\mu = 0$ imply that $M^{\mu\nu} = \chi_m F^{\mu\nu} $ and $H^{\mu\nu} = (1-\chi_m) F^{\mu\nu} $. This, in turn, leads to the relations
\beq
\Hvc = (1-\chi_m) \Bvc, \quad \Dvc = (1-\chi_m) \Evc.
\label{matrel}
\eeq
We note that one usually introduces the relation $\Dvc = (1+\chi_e) \Evc$ with $\chi_e$ being the electric susceptibility. In the case of ideal MHD, the relations between electric-type of fields follow from the relations between magnetic-type of fields that hold in LRF. Therefore, the factor $1+\chi_e$ is related to $1-\chi_m$. For constant $\chi_m$ we have $\mathbf{\nabla}\cdot \Dvc =   (1-\chi_m) \mathbf{\nabla}\cdot \Evc$
and $\mathbf{\nabla}\times \Hvc = (1-\chi_m) \mathbf{\nabla}\times \Bvc$. 

\subsubsection{Energy-momentum conservation}

With the assumptions made in this section it is useful to introduce a new notation for the sum of interaction and field parts of the energy-momentum tensor~\rfn{eq:TmnT}, 
\beq
T^{\mu\nu}_\Delta =  T^{\mu\nu}_\textrm{int} + T^{\mu\nu}_\textrm{em} =
 -H^{\mu\tau}F^\nu_{~\tau} + \frac{1}{4}g^{\mu\nu}F^{\tau\sigma}F_{\tau\sigma}. 
\label{eq:TmnD}
\eeq
Then, using the Bianchi identity, $\p_\alpha F_{\beta \gamma}+\p_\beta F_{\gamma \alpha}+\p_\gamma F_{\alpha \beta}=0$,  and the inhomogeneous Maxwell equations we find
\beq
\p_\mu T^{\mu\nu}_\Delta = -F^\nu_{~\tau} J^\tau + M^{\mu\tau} \p_\mu F^\nu_{~\tau} = -F^\nu_{~\tau} J^\tau + M \p^\nu B.
\eeq
This expression should be used in the conservation laws \rfn{eq:TmnC}, which can be written now as
\be
\partial_{\mu}T^{\mu\nu}_{\rm m}  + \p_\mu T^{\mu\nu}_\Delta= 0\, .
\label{eq:TmnCn}
\ee

\subsection{Equation of state}
\label{sec:EOS}

In this work, following Ref.~\cite{Pu:2016ayh}, we consider a simple equation of state of the form
\beq
\varepsilon(T,\mu_B,B) = \varepsilon_0(T,\mu_B) -\frac{1}{2} \chi_m B^2,
\eeq
\beq
p(T,\mu_B,B) = p_0(T,\mu_B) + \frac{1}{2} \chi_m B^2.
\eeq
The energy density $ \varepsilon_0(T,\mu_B)$ and pressure $p_0(T,\mu_B)$ describe equilibrium properties of matter without magnetic field. In this work, in addition to the temperature $T$ we introduce the baryon chemical potential $\mu_B$ as an independent thermodynamic variable. In this case, the basic thermodynamic identities are
\beq
\varepsilon + p = \varepsilon_0 + p_0 = T s_0 + n_B \mu_B,
\label{eq:thid1}
\eeq
\beq
dp = s_0 dT + n_B d\mu_B + M dB, \quad d\varepsilon = T ds_0 + \mu_B dn_B - M dB,
\label{eq:thid2}
\eeq
where $s_0$ is the entropy density and $n_B$ is the baryon number density. The conservation of the latter is expressed by the equation
\beq
\partial_\mu N_B^\mu = \partial_\mu (n_B U^\mu) = 0.
\label{eq:NBC}
\eeq

Below we use two forms of the equation of state defined by the two expressions for the pressure~\cite{Florkowski:2010zz},
\beq
p_0(T,\mu_B) &=& \frac{g}{\pi^2} \cosh(\mu_B/T) T^2 m^2 K_2(m/T), \quad (\hbox{EOS I}) \label{eos1} \\
p_0(T,\mu_B) &=& \frac{2g}{\pi^2} \cosh(\mu_B/T) T^4. \quad \hspace{2.25cm} (\hbox{EOS II}) \label{eos2}
\eeq
Here $g$ is the internal degeneracy factor and $K_2(x)$ is the modified Bessel function. Equation~\rfn{eos1} describes a system of classical (Boltzmann) particles and antiparticles with the mass $m$, while \rf{eos2} describes the same system in the limit $m \to 0$. Using equations of state \rfn{eos1} or \rfn{eos2}, one obtains the system's entropy and baryon number density from the thermodynamic relations
\beq
s_0 = \left( \frac{\p p_0}{\p T} \right)_{\mu_B} =  \left( \frac{\p p}{\p T} \right)_{\mu_B, B}, 
\quad n_B = \left( \frac{\p p_0}{\p \mu_B} \right)_T = \left( \frac{\p p}{\p \mu_B} \right)_{T, B} ,
\label{eq:thid3}
\eeq
which follow directly from Eqs.~\rfn{eq:thid1} and \rfn{eq:thid2}. The baryon number densities in these two cases are given by the formulas
\beq
n_B(T,\mu_B) &=& \frac{g}{\pi^2} \sinh(\mu_B/T) T m^2 K_2(m/T), \quad (\hbox{EOS I}) \label{eos1n} \\
n_B(T,\mu_B) &=& \frac{2g}{\pi^2} \sinh(\mu_B/T) T^3. \quad \hspace{2.125cm} (\hbox{EOS II}) \label{eos2n}
\eeq
We use these expressions below in Sec.~\ref{sec:geq}. To complete thermodynamic relations we add the equation specifying magnetization $M$ as a derivative of pressure with respect to magnetic field $B$,
\beq
M = \left( \frac{\p p}{\p B} \right)_{T,\mu_B} = \chi_m B.
\label{eq:thid4}
\eeq

\subsection{idMHD - summary}
\label{sec:idMHD-sum}
We can now summarize all the equations of idMHD introduced one by one in the previous sections. Using thermodynamic relations \rfn{eq:thid2}, they may be written as the following sequence of expressions:
\beq
\mathbf{\nabla}\cdot \Bvc&=&0, 
\label{eq:BGauss} \\     
\Evc + \mathbf{v} \times \Bvc&=&0, 
\label{eq:EOhm1} \\
\Evc \cdot\mathbf{v}&=&0, 
\label{eq:EOhm2} \\
\mathbf{\nabla}\times \Evc&=&-\frac{\partial \Bvc}{\partial{t}}, 
\label{eq:Faraday} \\
(1-\chi_m) \mathbf{\nabla}\cdot \Evc &=& \rho, 
\label{eq:rho} \\
(1-\chi_m) \mathbf{\nabla}\times \Bvc &=& \Jvc+ (1-\chi_m) \frac{\partial \Evc }{\partial{t}},  
\label{eq:current} \\
U^\mu \partial_{\mu} \varepsilon_0 &=&- \lt(\varepsilon_0 + p_0\rt) \partial_{\mu} U^\mu, 
\label{eq:econs} \\
\partial_{\mu}\lt(n_B U^{\mu}\rt)&=&0,  \label{eq:bcons} \\   
\lt(\varepsilon_0 +p_0 \rt) \Av&=&-\mathbf{\nabla} p_0 + \Evc \rho +\Jvc \times \Bvc.
 \label{eq:mcons}
\eeq
The reason for rewriting the idMHD equations in the form given above is that it naturally describes their structure and indicates their overall mathematical consistency. It also gives the hints how they can be solved~\footnote{The method presented in this section may be not the best procedure from the numerical point of view.} and approximated. 

Equation~\rfn{eq:econs} has been obtained from the longitudinal part of the energy-momentum conservation \rfn{eq:TmnCn} and the homogeneous Maxwell equations. Using the thermodynamic relations \rfn{eq:thid3} as well as the baryon number conservation \rfn{eq:bcons}   one obtains from~\rf{eq:econs}, as expected, the entropy conservation
\beq
\partial_{\mu}\lt(s_0 U^{\mu}\rt)&=&0 \label{scons} .
\eeq
Equation~\rfn{eq:mcons} follows from the transverse part of the energy-momentum conservation \rfn{eq:TmnCn} and represents the relativistic Euler equation for MHD.

The initial conditions for idMHD consist of two vector and two scalar functions of space: $\Bvc(t_0, \xv)$, $\vv(t_0,\xv)$, $T(t_0,\xv)$ and $\mu_B(t_0,\xv)$. The Gauss equation for the magnetic field implies that the initial condition for $\Bvc(t_0, \xv)$ is not arbitrary but has to fulfil \rf{eq:BGauss}. Using the initial condition for the hydrodynamic flow, $\vv(t_0,\xv)$, in \rf{eq:EOhm1} we obtain the initial electric field $\Evc(t_0, \xv)$. Then, \rf{eq:EOhm2} is automatically satisfied, while \rf{eq:Faraday} allows us to obtain the magnetic field at the next time step, which we denote as $\Bvc(t_0+\Delta t, \xv)$.  If \rf{eq:BGauss} is fulfilled at $t=t_0$, from \rf{eq:Faraday} one concludes that it is fulfilled at later times. From \rf{eq:rho} we find the initial charge density $\rho(t_0,\xv)$, while~\rfn{eq:current} defines the initial current $\Jvc(t_0,\xv)$ in terms of the gradient of the EM fields and the fluid acceleration (note that using \rf{eq:EOhm1} the time derivative of the electric field at $t=t_0$ can be expressed by the time derivative of the magnetic field and acceleration, $\p \Evc/\p t  + (\p \vv/\p t) \times \Bvc + \vv \times (\p \Bvc/\p t)=0$). The last three equations in the group \rfn{eq:BGauss}--\rfn{eq:mcons} express the energy, baryon number, and momentum conservation in the system. They may be used to determine the functions $\vv(t_0+\Delta t,\xv)$, $T(t_0+\Delta t,\xv)$ and $\mu_B(t_0+\Delta t,\xv)$. Repeating iteratively the steps described above, we are able to determine the space-time evolution of the fields $\Bvc(t, \xv)$, $\vv(t,\xv)$, $T(t,\xv)$ and $\mu_B(t,\xv)$.

\subsection{Standard MHD (stMHD)}
\label{sec:stMHD}
In the most popular approach to MHD one neglects the charge density and displacement vector in~Eqs.~\rfn{eq:BGauss}--\rfn{eq:mcons}. This is justified if there is a certain reference frame where the physical conditions are such that the terms with $\Jvc$ dominate over the terms $(1-\chi_m) \partial \Evc/\partial{t}$ and $\Evc \rho$ (on the right-hand sides of Eqs.~\rfn{eq:current} and \rfn{eq:mcons}). The charge density can be usually neglected if the system consists of two components with opposite charges, like in the case of ``ordinary'' plasma formed from protons and electrons.  Using these assumptions we obtain the following set of equations:
\beq
\mathbf{\nabla}\cdot \Bvc&=&0, \label{BGauss} \\     
\Evc + \mathbf{v} \times \Bvc&=&0, \label{EOhm1} \\
\mathbf{\nabla}\times \Evc&=&-\frac{\partial \Bvc}{\partial{t}}, \label{Faraday} \\
(1-\chi_m) \mathbf{\nabla}\times \Bvc &=& \Jvc,  
\label{current} \\
U^\mu \partial_{\mu} \varepsilon_0 + \lt(\varepsilon_0 + p_0\rt) \partial_{\mu} U^\mu&=& 0, \label{econs} \\
\partial_{\mu}\lt(n_B U^{\mu}\rt)&=&0,  \label{bcons} \\   
\lt(\varepsilon_0 +p_0\rt) \Av&=&-\mathbf{\nabla} p_0 +\Jvc \times \Bvc.
 \label{mcons}
\eeq
It should be emphasized that Eqs.~\rfn{BGauss}--\rfn{mcons},  in contrast to Eqs.~\rfn{eq:BGauss}--\rfn{eq:mcons}, are no longer covariant with respect to Lorentz transformations.  They hold in a reference frame where the assumptions specified above are justified. Nevertheless, if the solutions of Eqs.~\rfn{BGauss}--\rfn{mcons} are found, they can be transformed to any other frame. We also note that neglecting the term $(1-\chi_m) \partial \Evc/\partial{t}$ facilitates substantially the mathematical treatment of the MHD equations. 

\subsection{Manifestly covariant MHD (cvMHD)}
\label{sec:cvMHD}

In this section we finally introduce the equations of covariant relativistic MHD. Using the parametrization \rfn{eq:field_strength_tensor} for the field tensor $F_{\mu\nu}$ and the condition \rfn{ohm's_law}, we can rewrite the energy-momentum tensor~\rfn{eq:TmnD} as~\cite{Gedalin,PhysRevE.47.4354,PhysRevE.51.4901,PhysRevLett.76.3340,Florkowski:2009sw,Roy:2015kma,Pu:2016ayh,Inghirami:2016iru}
\beq
T^{\mu\nu}_\Delta =  \left( B^2 - M B \right) U^\mu U^\nu + \left( M B - \frac{1}{2} B^2 \right) g^{\mu\nu}
+ \left( M B - B^2 \right) b^\mu b^\nu.
\label{eq:TmnDb}
\eeq
Using this expression in the conservation laws for energy and momentum,  the MHD equations can be cast into the following compact form:
\beq
U^\mu \partial_{\mu} \varepsilon_0 + \lt(\varepsilon_0 + p_0\rt) \partial_{\mu} U^\mu&=& 0, 
\label{eq:econs2} \\
\left[ \varepsilon + p+ (1-\chi_m) B^2 \vphantom{\frac{1}{2}}  \right]  A^\alpha
-\Delta^{\alpha\mu}\partial_{\mu} \left[ p+ \left( \frac{1}{2}-\chi_m \right) B^2 \right] \nn\\
- (1-\chi_m) \left[b^{\alpha}   b^{\mu} \p_\mu B^2 
+ B^2  (   \Delta^{\alpha \rho} b^{\mu}\partial_{\mu}b_{\rho}
+b^{\alpha} \partial_{\mu}b^{\mu}) \vphantom{\frac{1}{2}} \right]&=&0,
\label{eq:mcons2} \\
\partial_{\mu}\lt(n_B U^{\mu}\rt)&=&0,
\label{eq:bcons2}  \\
\partial_{\nu} \left[ B (b^{\mu} U^{\nu}-b^{\nu} U^{\mu}) \right]&=&0.
\label{eq:maxwellU}
\eeq
Similarly as in the previous cases, \rf{eq:econs2} combined with \rf{eq:bcons2} yield the entropy conservation~\rfn{scons}.
We note that Eqs.~\rfn{eq:maxwellU} are equivalent to Eqs.~\rfn{eq:maxwell_dual}, if the condition \rfn{ohm's_law} is used. We also note that there are only three independent equations in~\rfn{eq:mcons2}, hence Eqs.~\rfn{eq:econs2}--\rfn{eq:maxwellU} are nine independent equations for eight unknowns: $T$, $\mu_B$, $\vv$, and $\Bv$. This looks like an overdetermined system but, as the matter of fact, Eqs.~\rfn{eq:maxwellU} are the uniform Maxwell equations containing the Gauss law for the magnetic field. We know that if the magnetic field is divergence-free at the initial time it remains divergence-free at later times, hence, by imposing the proper initial conditions only three out of four equations in \rfn{eq:maxwellU} are truly independent. 

\begin{figure}[t]
\includegraphics[angle=0,width=0.6\textwidth]{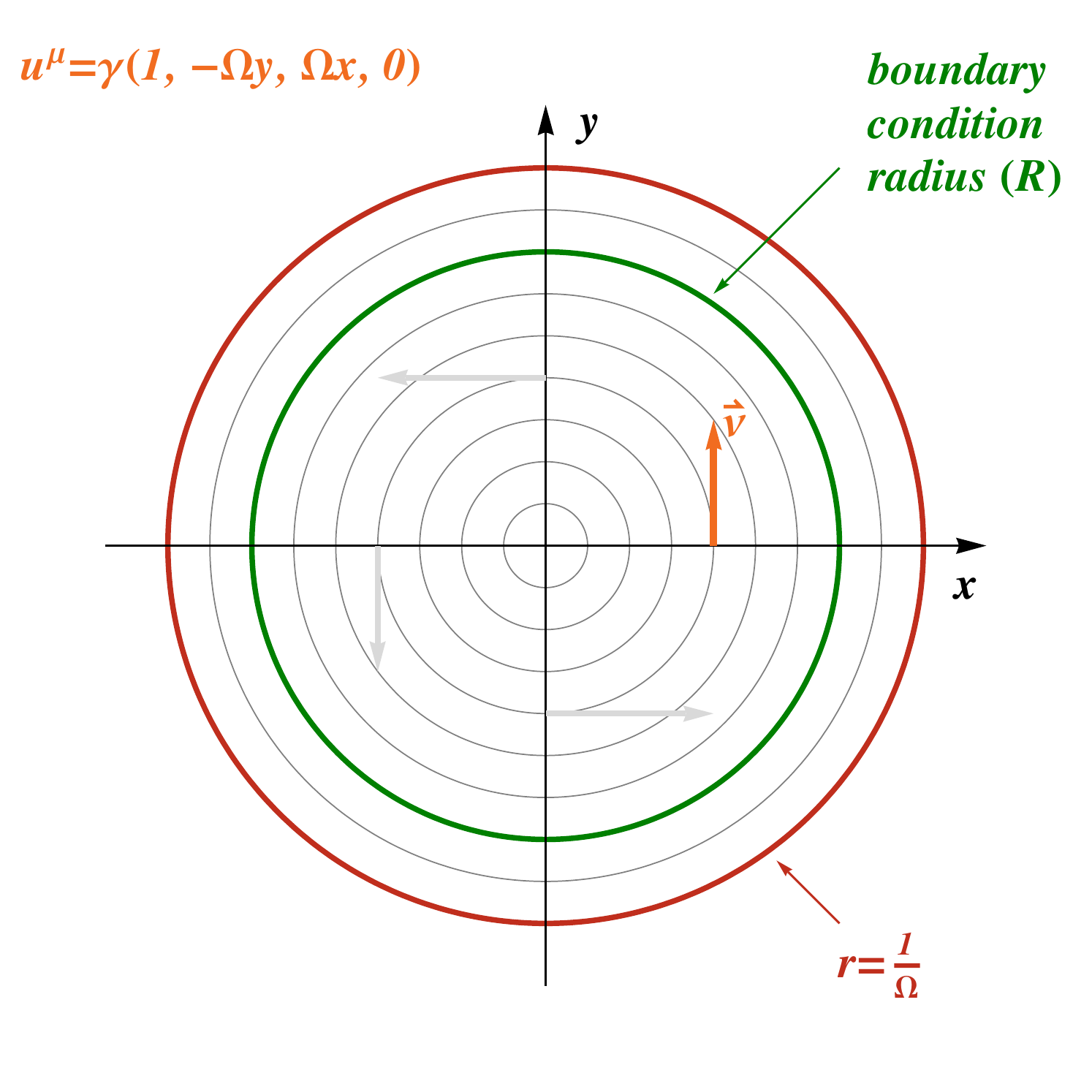} 
\caption{(Color online)  Vortex configuration described in Sec.~\ref{sec:vortex}.}.
\label{fig:vortex1}
\end{figure}
%

\section{Vortex-like solutions}
\label{sec:vortex}
In this section we present the vortex-like solutions of the cvMHD equations introduced in Sec.~\ref{sec:cvMHD}. They become automatically solutions of the idMHD equations presented  in Sec.~\ref{sec:idMHD-sum}. Following \cite{Florkowski:2017ruc}, we assume that the flow four-vector $U^\mu$ has the form
\be
 U^{\mu}=\gamma (1,-\Omega y,\Omega x,0),
 \label{eq:flow_velocity}
 \ee
 where $\Omega$ is a constant, $\gamma=(1-\Omega^2r^2)^{-1/2}$, and $r=\sqrt{x^2+y^2}$ is the distance from the vortex center in the transverse plane. Note that, due to limited speed of light, the flow profile is constrained to a cylinder of radius $R<1/\Omega$, see Fig.~\ref{fig:vortex1}. The larger is the value of $R$, the larger are the values of the $\gamma$ factor on the boundary where $r=R$. 
 
 Furthermore, we assume that the magnetic field is perpendicular to the vortex plane, 
\be
B^\mu = (0,0,0,B(r)), \quad b^\mu = (0,0,0,1).
\label{eq:Br}
\ee
Using \rftwo{eq:A}{eq:flow_velocity} we find the following relations involving the fluid four-velocity and three-acceleration
\beq
U^\mu \p_\mu &=& -\gamma \Omega \left( y \frac{\p}{\p x} - x \frac{\p}{\p y} \right), \label{eq:ddtau} \\
\p_\mu U^\mu &=& 0 \label{eq:expsc}, \\
\Av &=& - \Omega^2 \gamma^2 (x,y,0) \label{eq:A2}.
\eeq
The three-vector $\Av$ points towards the centre of the vortex, describing the centripetal acceleration.

We look for stationary (independent of time) solution and assume that all thermodynamic functions depend only on the variable $r = \sqrt{x^2+y^2}$. In this case \rftwo{eq:econs2}{eq:bcons2} are automatically fulfilled. Moreover, since $b^\mu$ is a constant vector and 
the function $B(r)$ is independent of the coordinate $z$,  \rf{eq:maxwellU} is also satisfied and we are left with only one non-trivial equation that reads
\beq
\Omega^2 \gamma^2 \left[ \varepsilon_0 + p_0+ (1-\chi_m) B^2 \vphantom{\frac{1}{2}}    \right]  
&=&  \frac{d}{r \, dr} \left[ p_0 + \frac{1}{2} \left(1-\chi_m \right) B^2 \right].
\label{eq:radial} 
\eeq
Using thermodynamic identities \rfn{eq:thid1} and \rfn{eq:thid2} we can rewrite this equation as
\beq
s_0 \left(\Omega^2 \gamma^2 T - \frac{dT}{rdr} \right)
+ n_B \left(\Omega^2 \gamma^2 \mu_B - \frac{d\mu_B}{rdr} \right)
+  \left(1-\chi_m \right) B \left(\Omega^2 \gamma^2 B - \frac{dB}{rdr} \right) = 0.
\label{eq:Radial}
\eeq
Since \rf{eq:Radial} is a single equation involving three functions of $r$, it allows for different solutions. They are defined by specific boundary conditions and possible additional constraints. 

\medskip
Before we discuss some of the solutions of \rf{eq:Radial}, it is useful to discuss direct consequences of our assumptions \rfn{eq:flow_velocity} and \rfn{eq:Br}  for other physical quantities, such as the electric field, electric charge, and electric three-current. For the field configuration \rfn{eq:Br}, the magnetic field has the form $\Bvc = (0,0, \gamma B(r))$, which automatically satisfies the Gauss law $\nabla \cdot \Bvc = 0$. The electric field is given by the expression $\Evc = -\Omega \gamma B(r) (x,y,0)$, which is consistent with the Ohm law \rfn{eq:EOhm1}, i.e., $\Evc = - \vv \times \Bvc$. The inhomogeneous Maxwell equations define the electric charge
\beq
\rho = (1-\chi_m) \nabla \cdot \Evc =  \Omega_m \gamma B(r)  \left[ \gamma^2 +1 + \frac{d\ln B(r)}{d\ln r} \right]
\label{eq:rhov}
\eeq
as well as the electric three-current
\beq
\Jvc = (1-\chi_m) \nabla \times \Bvc 
&=&  \Omega_m \gamma B(r) \left[  \gamma^2  + \frac{1}{\Omega^2 r^2}  \frac{d\ln B(r)}{  d \ln r}\right] \vv,
\label{eq:jv}
\eeq
where we have introduced the notation $\Omega_m = -(1-\chi_m)  \Omega$. We see that the electric current is proportional to the three-velocity of the fluid. Interestingly, the proportionality coefficient is not necessarily equal to the electric charge $\rho$, as naively expected. This means that the electric current is generally not represented by the convective part in \rf{eq:Jsum}.

\subsection{Global equilibrium with rotation}
\label{sec:geq}
In a series of publications Becattini and collaborators showed that global equilibrium corresponds to the case where the field $\beta_\mu = U_\mu/T$ is the Killing vector~\cite{Becattini:2009wh}. In our case, this implies that
\beq
T(r) = T_{\rm c} \gamma(r),  \label{eq:Tgeq}
\eeq
where $T_c$ is the central temperature. In global equilibrium we also find that $\mu_B/T$=const., hence, the chemical potential has the same form as the temperature, namely
\beq
\mu_B(r) = \mu_B^{\rm c} \gamma(r), \label{eq:muBgeq}
\eeq
where $\mu_B^{\rm c}$ is another constant. The forms \rfn{eq:Tgeq} and \rfn{eq:muBgeq} imply that the first two terms on the left-hand side of \rf{eq:Radial} vanish. This, in turn, implies that 
\beq
B(r) = B_{\rm c} \gamma(r),
\label{eq:Bvgeq}
\eeq
where $B_{\rm c}$ is a constant defining the magnetic field at the vortex center.

In the case described by Eqs.~\rfn{eq:Tgeq}, \rfn{eq:muBgeq}, and \rfn{eq:Bvgeq},   we find
\beq
\rho = 2\Omega_m  \gamma^3 B(r),
\eeq
and 
\beq
\Jvc = \rho \vv.
\eeq
Thus, using the notation $\rho = {\bar \rho} \gamma$, where $ {\bar \rho} $ can be interpreted as the charge density measured in LRF, we can rewrite the EM four-current in the relativistic covariant way as
\beq
J^\mu = {\bar \rho} \, U^\mu, \qquad   {\bar \rho}  = 2\Omega_m B_{\rm c}   \gamma^3.
\label{Jmu}
\eeq
It is interesting to notice that using \rftwo{eq:rhov}{eq:jv}  and demanding the equality $\Jvc = \rho \vv$, one gets the constraint equation for the function $B(r)$,
\beq
1 + \frac{d\ln B(r)}{d\ln r}  = \frac{1}{\Omega^2 r^2}  \frac{d\ln B(r)}{  d \ln r},
\eeq
which has the solution given exactly by \rf{eq:Bvgeq}.

We observe that the charge density given by \rf{Jmu} can be very large, especially if the size of the system $R$ approaches the value of $1/\Omega$ and the $\gamma$  factor suddenly increases at the system boundary. Since we deal with a vortex, there is no boost transformation that leads to a different reference frame where $\rho$ can be globally reduced. Such a situation may contradict the usual assumptions of MHD which is applicable most often to quasi neutral systems. In our case, to avoid this problem one can restrict oneself to small vortices, with the size $R$ much smaller than $1/\Omega$, and choose the parameters $\Omega_m$ and $B_{\rm c}$ to be sufficiently small. In the general case of solving the cvMHD equations, one cannot control the charge density and the results of the calculations should be checked a posteriori if they obey the standard MHD criteria. 

\medskip
Another interesting observation can be made by comparison of the electric charge density $ {\bar \rho} $ with the baryon density defined by Eqs.~\rfn{eos1n} and \rfn{eos2n}. Using Eqs.~\rfn{eq:Tgeq} and \rfn{eq:muBgeq} in Eqs.~\rfn{eos1n} and \rfn{eos2n} we obtain
\beq
n_B &=& \frac{g}{\pi^2} \sinh(\mu^c_B/T_c) T_c \gamma m^2 K_2(m/(T_c \gamma)), \quad (\hbox{for EOS I}) \label{eos1nr} \\
n_B &=& \frac{2g}{\pi^2} \sinh(\mu^c_B/T_c) T_c^3 \gamma^3. \quad \hspace{2.7cm} (\hbox{for EOS II}) \label{eos2nr}
\eeq
Thus, we conclude that only for the massless case the baryon number and electric charge densities are proportional to each other, as expected for a system consisting of one type of particles (and their antiparticles)~\footnote{We note that conflicts between equation of state and vortical motion were found in the past in the context of the Bjorken flow~\cite{Bialas:1984gw,Dyrek:1984xz}.}
. Clearly, a simple kinetic-theory interpretation that the flow four-vector and all other conserved currents are expressed as moments of a single phase-space distribution function multiplied by the corresponding charges is not valid.

\subsection{Quasi-neutral vortex}
\label{sec:zerocharge}

In this section we analyze a possibility of having a vortex with a constant charge density. We do not assume any relations of the form \rfn{eq:Tgeq}, \rfn{eq:muBgeq}, or \rfn{eq:Bvgeq} but start with \rf{eq:rhov} and demand that its right-hand side is equal to a constant charge density $\rho_0$, 
\beq
\Omega_m \gamma B(r)  \left[ \gamma^2 +1 + \frac{d\ln B(r)}{d\ln r} \right] = \rho_0.
\label{eq:rho0}
\eeq
Equation \rfn{eq:rho0} is a differential equation for the function $B(r)$ which has the solution
\beq
B(r) = \frac{ \rho_0}{2 \Omega_m} \sqrt{1-\Omega^2 r^2} + C \, \frac{ \sqrt{1-\Omega^2 r^2}}{r^2} ,
\label{Br01}
\eeq
where $C$ is an integration constant. In order to have a regular solution we set $C$ equal to zero and keep only the first term in \rfn{Br01},
\beq
B(r) = \frac{ \rho_0}{2 \Omega_m} \sqrt{1-\Omega^2 r^2} = \frac{ \rho_0}{2 \Omega_m \gamma} .
\label{Br02}
\eeq
In this case, using \rf{eq:jv} we find that the electric three-vector vanishes, $\Jvc = 0$. This is in agreement with the fact that the magnetic field is constant, $\Bvc =  \rho_0/(2 \Omega_m)(0,0,1)$, and has vanishing rotation.

The case discussed in this section is more difficult for kinetic-theory interpretation than the previous one, since the electric current is not given by the convective part. One may argue that the induction part is finite in this case, hence, the term $J^\mu_{\rm ind} = \sigma^{\mu \nu} E_\nu$ compensates the difference between $J^\mu$ and $J^\mu_{\rm con} = {\bar \rho} U^\mu$. However, this approach is not quite satisfactory, since implies a random use of the term $\sigma^{\mu \nu} E_\nu$, that depends on a particular cvMHD solution rather than the system physical properties.

The knowledge of the function $B(r)$ allows us to determine the functions $T(r)$ and $\mu_B(r)$ from \rf{eq:Radial}, which in this case takes the form
\beq
s_0 \left(\Omega^2 \gamma^2 T - \frac{dT}{rdr} \right)
+ n_B \left(\Omega^2 \gamma^2 \mu_B - \frac{d\mu_B}{rdr} \right)
+  \left(1-\chi_m \right) \frac{\Omega^2 \rho_0^2}{2 \Omega_m^2} = 0.
\label{eq:Radial0}
\eeq
As this is one equation for two unknown functions, various forms of $T(r)$ and $\mu_B(r)$ are allowed in this case, which depend on the boundary conditions and the choice of EOS. For sufficiently small values of the parameter ${\rho_0}$, Eqs.~\rfn{eq:Tgeq} and \rfn{eq:muBgeq} are approximate solutions of \rf{eq:Radial0}.

In the similar way to the case described in this section one can treat other cases with different dependence of the charge density on the transverse distance $r$.

\section{Conclusion}

In this paper we have analyzed a manifestly covariant formulation of ideal relativistic magnetohydrodynamics (cvMHD)  and compared it with other frameworks (idMHD and stMHD). We have demonstrated that the covariant approach allows for vortex-like solutions that can be treated as generalizations of the global equilibrium states with rotation.

Our solutions indicate that both large charge densities and large electric currents can be present in a system described by the cvMHD equations. This is due to the implicit treatment of the inhomogeneous Maxwell equations in this framework, and no direct control of the relative magnitude of the ``density'' and ``current'' terms. 

Moreover, we have found that the flow four-vector $U^\mu$, the electric charge current $J^\mu$, and the baryon number current $N^\mu_B$ might be not related to each other if obtained within cvMHD. This poses interesting questions concerning kinetic-theory interpretations of the results obtained with the relativistic covariant  version of MHD.  Clearly, the knowledge of $U^\mu$ is not sufficient to determine the conserved currents.

Our results can be useful for making physics interpretation of the cvMHD calculations done in astrophysics and heavy-ion collisions.

{\bf Acknowledgements:} W.F. thanks Stanis\l aw Mr\'owczy\'nski for useful discussions. This work was supported in part by the Polish National Science Center Grant No. 2016/23/B/ST2/00717.


\addcontentsline{toc}{chapter}{Bibliography}
\bibliography{hydro_review}{}
\bibliographystyle{utphys}

\end{document}